\documentclass[osajnl2,twocolumn,showpacs]{revtex4}

\usepackage[draft]{hyperref}
\usepackage{amsmath}
\usepackage{calc}
\usepackage{amsfonts}
\usepackage{amssymb}
\usepackage{graphicx}
\usepackage{bm}
\usepackage{color}
\usepackage{epsfig}
\usepackage{ifthen}

\def\J{\mathbf{J}}

\begin{document}

\title{Highly efficient broadband conversion of light polarization by composite retarders}

\author{Svetoslav S. Ivanov$^{1,*}$, Andon A. Rangelov$^{1}$, and Nikolay V. Vitanov$^{1}$}
\address{
$^1$ Department of Physics, Sofia University, James Bourchier 5 blvd., 1164 Sofia, Bulgaria\\
$^*$Corresponding author: sivanov@phys.uni-sofia.bg }

\author{Thorsten Peters$^{2}$ and Thomas Halfmann$^{2}$}
\address{
$^2$ Institut f\"ur Angewandte Physik, Technische Universit\"at Darmstadt, Hochschulstr. 6, D-64289 Darmstadt, Germany}

\begin{abstract}
Driving on an analogy with the technique of composite pulses in quantum physics, we propose highly efficient broadband polarization converters composed of sequences of ordinary retarders rotated at specific angles with respect to their fast-polarization axes.
\end{abstract}


\maketitle
\section{Introduction}

Broadband (achromatic) polarization retarders of light have been a subject of significant interest in optics for several decades \cite{Wolf,Azzam,Goldstein}.
Such retarders are assembled by combining two or more ordinary wave plates, either of the same or different material.
One of the first documented proposals was by West and  Makas \cite{West49} who described achromatic combinations of plates  having  different dispersions of birefringence.
Achromatic retarders composed of wave plates of the same material but different thicknesses were proposed by
 Destriau  and Prouteau \cite{Destriau49} for two birefringent plates
 and Pancharatnam for three plates, with which he constructed half-wave \cite{Pancharatnam55a} and quarter-wave \cite{Pancharatnam55b} retarders.
Later Harris and co-workers proposed achromatic quarter-wave plates with 6 \cite{Harris64} and 10 identical zero-order quarter-wave plates \cite{McIntyre68}.

These early studies used either the Stokes vector or the Jones vector description of light polarization.
The equations of motion of these vectors which describe the polarization change when light passes through an anisotropic optical medium \cite{Wolf,Azzam,Goldstein}
 are identical to basic equations in quantum mechanics:
 the equation for the Jones vector in a medium with zero polarization-dependent loss is identical to the Schr\"{o}dinger equation \cite{Kubo78},
 while the Stokes vector obeys an equation of the same form as the Bloch equation \cite{Kubo80}.
This analogy was used to link the dynamics of light polarization to the dynamics of two-state quantum systems,
 such as a spin-1/2 particle in a magnetic field or a two-level atom in a laser field \cite{Kuratsuji1,Kuratsuji2,Rangelov,Botet1,Botet2};
 this analogy has paved the ground for the linear optics implementation of quantum computing \cite{KLM}.

The composite achromatic retarders discussed above \cite{West49,Destriau49,Pancharatnam55a,Pancharatnam55b,Harris64,McIntyre68} are the precursor
 of the composite pulse sequences discovered later, apparently independently, in nuclear magnetic resonance (NMR) \cite{CompositeNMR1,CompositeNMR2,Wimperis}.
Composite pulse sequences are widely used in NMR to manipulate spins with high fidelity and robustness to parameter variations;
 they have a significant potential in quantum optics \cite{IvanovLocal,Torosov,Torosov2} and quantum information processing \cite{Haffner,IvanovGate} as well.
Ardavan \cite{Ardavan} has recently recognized this analogy and proposed a broadband composite linear retarder based on some well-known composite pulses in NMR \cite{CompositeNMR1,CompositeNMR2,Wimperis}.

In this paper, we use the analogy between the polarization Jones vector and the quantum state vector to propose \emph{arbitrarily accurate} broadband polarization retarders.
To this end, we design novel phase-stabilized broadband (BB) composite sequences, which outperform the BB1 sequence \cite{Wimperis} used by Ardavan \cite{Ardavan}
 and the other achromatic retarders proposed earlier.
Our composite retarders promise to deliver very high polarization conversion fidelity in an arbitrarily broad range of wavelengths.

\section{Background}

Any polarization system can be viewed as composition of a retarder and a rotator \cite{Hurvitz}.
A rotation at angle $\theta$ in the polarization plane is described by the Jones matrix
\begin{equation}
\mathfrak{R}(\theta)=\left[
\begin{array}{cc}
\cos \theta & \sin \theta \\
-\sin \theta & \cos \theta%
\end{array}%
\right].
\end{equation}
A retarder increases the phase of the electric field by $\varphi /2$ along the fast axis and retards it by $-\varphi /2$ along the slow axis,
which can be expressed in the horizontal-vertical (HV) basis by the Jones matrix
\begin{equation}
\mathfrak{J}(\varphi)=\left[
\begin{array}{cc}
e^{i\varphi/2} & 0 \\
0 & e^{-i\varphi/2}%
\end{array}%
\right],
\end{equation}%
where $\varphi = 2\pi L ( n_{f}-n_{s})/\lambda$, with $\lambda$ being the vacuum wavelength, $n_{f}$ and $n_{s}$ the refractive indices along the fast and slow axes, respectively,
 and $L$ the thickness of the plate.
The most common type of retarders are the half-wave plates ($\varphi =\pi$) and quarter-wave plates ($\varphi =\pi /2$).
Since the performance of such retarders, i.e. the phase shift $\varphi$, depends strongly on the thickness and the rotary power of the plate,
 the traditional wave plates are not broadband, as only a narrow range of wavelengths around $\lambda$ acquire the desired phase shift.
To this end, we use composite wave plates to create broadband retarders.

Let us now consider a single polarizing birefringent plate of phase shift $\varphi$ and let us introduce a system of HV polarization axes (HV basis),
which are rotated by an angle $\theta$ with respect to the slow and the fast axes of the plate.
The Jones matrix $\mathfrak{J}$ has the form
\begin{equation} \label{jones2}
\mathfrak{J}_{\theta}(\varphi) = \mathfrak{R}(-\theta) \mathfrak{J}(\varphi) \mathfrak{R}(\theta).
\end{equation}
In the left-right circular polarization (LR) basis this matrix obtains the form $\J_{\theta}(\varphi) = \mathbf{W}^{-1}\mathfrak{J}_{\theta}(\varphi)\mathbf{W}$, where $\mathbf{W}$ connects the HV and LR polarization bases,
\begin{equation}
\mathbf{W} = \tfrac{1}{\sqrt{2}} \left[\begin{array}{cc}
1 & 1 \\ -i & i
\end{array}\right].
\end{equation}
Explicitly, the Jones matrix in the LR basis is
\begin{equation}
\J_{\theta}(\varphi) = \left[\begin{array}{cc}
\cos \left( \varphi /2\right) & i\sin \left( \varphi /2\right) e^{2i\theta } \\
i\sin \left( \varphi /2\right) e^{-2i\theta } & \cos \left( \varphi /2\right)
\end{array}\right] .
\end{equation}
Half- and quarter-wave plates rotated by angle $\theta$, $(\lambda/2)_{\theta}$ and $(\lambda/4)_{\theta}$, are respectively described by $\J_{\theta}(\pi)$ and $\J_{\theta}(\pi/2)$.

\section{Composite polarization retarders}

Our objective is to construct retarders that are robust to variations in the phase shift $\varphi$ at selected value(s) of this shift.
Such retarders tolerate imperfect rotary power $\varphi /L$ and plate thickness $L$ and also operate over a broad range of wavelengths $\lambda$.
To this end, we replace the single retarder with a sequence of $N$ retarders, each with phase shift $\varphi _{k}$ and rotated by angle $\theta _{k}$, described in the LR basis by the Jones matrix (read from right to left)
\begin{equation}
\J^{\left( N\right) }=\J_{\theta _{N}}\left( \varphi _{N}\right) \J_{\theta_{N-1}}\left( \varphi _{N-1}\right) \cdots \J_{\theta _{1}}\left( \varphi_{1}\right) . \label{overall Jones matrix}
\end{equation}
The efficiency of the composite retarder is measured by the fidelity $F=\frac{1}{2}\text{Tr}\left( \J_0^{-1} \J^{(N)}\right)$ \cite{Ardavan},
 where $\J^{(N)}$ is the achieved and $\J_0$ the target Jones matrix.

\begin{figure}[t]
\centerline{\includegraphics[width=0.85\columnwidth,angle=0]{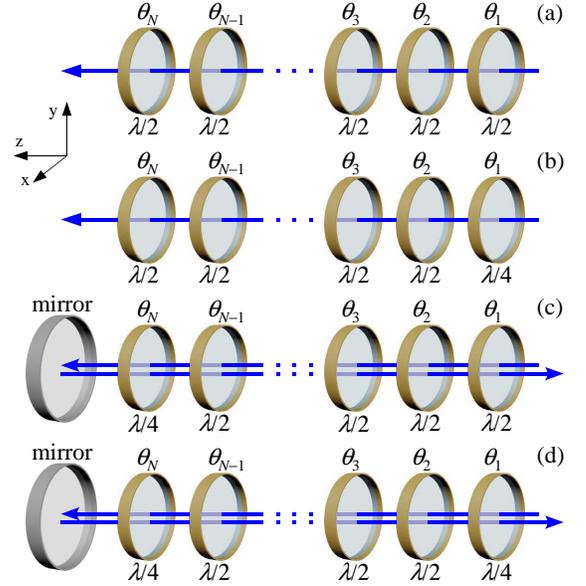}}
\caption{
(Color online) Implementations of different BB composite retarders
 composed of $N$ half- and quarter-wave plates, each rotated by angle $\theta_k$ $(k=1,2,\ldots, N)$ with respect to the fast axis, which is chosen to be along the $y$-axis.
 (a) BB half-wave retarder composed of $N$ half-wave plates;
 (b) BB quarter-wave retarder composed of $N-1$ half-wave plates and a quarter-wave plate;
 (c) BB half-wave retarder composed of $N-1$ half-wave plates, a quarter-wave plate, and a mirror;
 (d) BB quarter-wave retarder composed of $N-2$ half-wave plates, two quarter-wave plates, and a mirror.
}
\label{Fig1}
\end{figure}

First, we show how to construct a \emph{BB half-wave retarder}, which inverts light polarization with high fidelity in a broad range of phase shifts $\varphi$ around $\pi$.
Its Jones matrix in the LR basis reads (up to a global phase factor)
\begin{equation}
\J_{0}(\pi ) = \left[\begin{array}{cc}
0 & i \\ i & 0
\end{array}\right] . \label{half wave plate Jones matrix}
\end{equation}
We compose a sequence of an odd number $N$ of $\lambda/2$ plates ($\varphi_k=\varphi=\pi$), each rotated by $\theta_{k}$:
$(\lambda /2)_{\theta_N}(\lambda /2)_{\theta_{N-1}}\ldots(\lambda/2)_{\theta_1}$ (read from right to left, cf.~Fig.~\ref{Fig1}(a)).
The phase $\theta_{1}$ does not change the performance of our retarders. Therefore for convenience we set $\theta_{1}=0$ and we are left with $N-1$ relative rotation angles, which are treated as free parameters.

\begin{table}[t]
\begin{tabular}{|c|l|}
\hline
$N$ & Rotation angles $(\theta_{1}$; $\theta_{2}$; $\cdots$; $\theta_{N-1}$; $\theta_{N}$) \\ \hline
 & (a) half-wave retarders \\ \hline
5  & (0; 52.2; 336.7; 336.7; 52.2)\\
9  & (0; 158; 59.9; 45.5; 151.6; 174.4; 108.4; 0.5; 58.9)\\
13 & (0; 109.1; 175; 49; 91.6; 9.7; 172.8; 28.6; 127.3; \\
   &  131.2; 174.8; 76.1; 22.1)\\
\hline\hline
 & (b) quarter-wave retarders \\ \hline
4 & (0; 69.3; 318.6; 69.3) \\
5 & (0; 48.6; 325.8; 325.8; 48.6) \\
6 & (0; 116.6; 69.1; 175; 69.1; 116.6) \\
8 & (0; 104.8; 103.6; 32.5; 149.6; 52.6; 74.3; 137.6) \\
10& (0; 34.3; 97.1; 120.6; 142.7; 50.7; 8.9; 121.2; 64.9; 77) \\
\hline\hline
 & (c) half-wave retarders with a mirror\\ \hline
2 & (30; 150) \\
3 & (26.5; 55.1; 122.7) \\
4 & (22; 73.4; 171.6; 150.2) \\
5 & (9.4; 115.1; 154.7; 51; 3.8) \\
10& (110.3; 161.4; 18.8; 111.7; 96; 84.9; 136.2; 65.7; \\
  &  5.9; 67.1) \\
\hline\hline
  & (d) quarter-wave retarders with a mirror \\ \hline
4 & (45; 19.3; 113.6; 166) \\
5 & (45; 131.1; 169.3; 39.7; 115.5) \\
6 & (45; 130.3; 164.6; 25.1; 83; 162) \\
8 & (45; 141.5; 104.9; 127.2; 43.2; 65.9; 6.2; 112) \\
9 & (45; 152.7; 73.2; 30.1; 1.2; 144.4; 94.4; 29; 127.4) \\
10& (45; 126.6; 165.5; 145.1; 154.9; 64; 43; 25.3; \\
  &  87.1; 156.5) \\ \hline
\end{tabular}
\caption{Rotation angles $\theta_{k}$ (in degrees) for BB retarders with different number $N$ of constituent half-wave plates.
(a) half-wave BB composite retarders, cf. Fig.~\ref{Fig1}(a);
(b) quarter-wave BB composite retarders, cf. Fig.~\ref{Fig1}(b);
(c) half-wave BB composite retarders with a mirror, cf. Fig.~\ref{Fig1}(c);
(d) quarter-wave BB composite retarders with a mirror, cf. Fig.~\ref{Fig1}(d).
}
\label{Table}
\end{table}

\begin{figure}[t]
\centerline{\includegraphics[width=0.95\columnwidth]{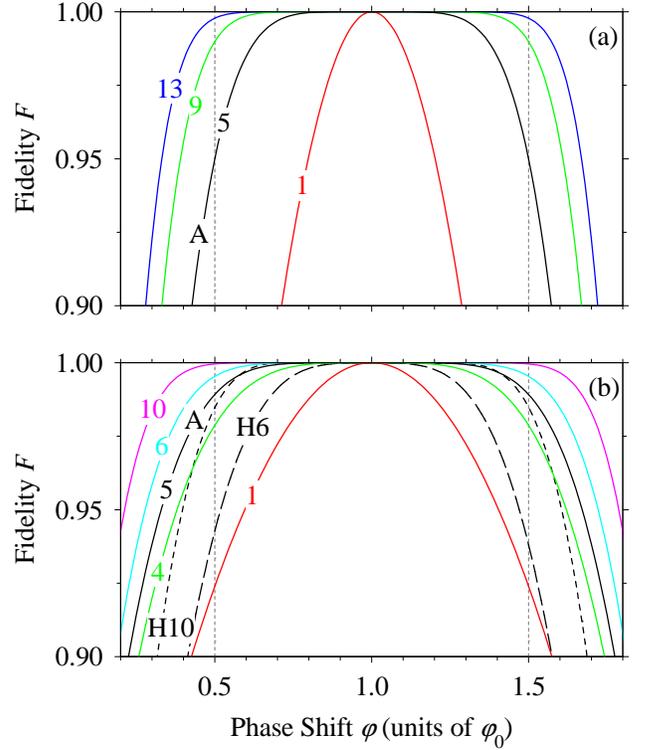}}
\caption{(Color online) Fidelity $F$ vs phase shift $\varphi$ for
 BB half-wave retarders (frame (a) with $\varphi_0=\pi$, cf.~Fig.~\ref{Fig1}(a)) and
 BB quarter-wave retarders (frame (b) with $\varphi_0=\pi/2$, cf.~Fig.~\ref{Fig1}(b)),
 for different number of constituent plates $N$.
The rotation angles are given in Table \ref{Table}.
The fidelities of the single-plate retarder and the BB1 retarder \cite{Ardavan} are shown with labels ``1'' and ``A'',
 while the fidelities of the 6- and 10-plate retarders of Harris and co-workers \cite{Harris64,McIntyre68} by dashed lines with labels ``H6'' and ``H10''. As one can clearly see, our retarders outperform the others for $N>5$.
}
\label{Fig2}
\end{figure}

Next, we obtain the composite Jones matrix \eqref{overall Jones matrix}
\begin{equation}
\J^{(N)}=\J_{\theta_N}(\pi)\J_{\theta_{N-1}}(\pi)\cdots \J_{\theta_1}(\pi)
\end{equation}
and set $\J^{(N)}=\J_0(\pi)$ at $\varphi=\pi$, which leaves us with $N-2$ independent angles $\theta_k$ to vary. We then nullify as many lowest order derivatives of $\J_{12}^{(N)}$ vs the phase shift $\varphi$ at $\varphi=\pi$ as possible.
We thus obtain a system of nonlinear algebraic equations for the $N-2$ rotation angles $\theta_k$.
For our composite retarder composed of $\lambda/2$ plates only, all odd-order derivatives vanish at $\varphi=\pi$;
hence, $N-2$ rotation angles allow us to nullify the first $N-2$ complex derivatives:
\begin{equation}
\left[\partial_{\varphi }^{k}\J_{12}^{\left( N\right) }\right] _{\varphi=\pi} = 0 \quad \left( k=1,2,...,N-2 \right)
\label{nullify11}
\end{equation}
as well as the real or imaginary part of the next non-zero derivative (of order $N-1$):
\begin{equation}
\text{Re}\left[\partial_{\varphi }^{N-1}\J_{12}^{\left( N\right) }\right] _{\varphi=\pi} = 0 \text{ or } \text{Im}\left[\partial_{\varphi }^{N-1}\J_{12}^{\left( N\right) }\right] _{\varphi=\pi} = 0
\label{nullify12}
\end{equation}
Solutions to Eqs. \eqref{nullify11} and \eqref{nullify12} provide BB half-wave retarders.
Longer retarders, of larger number $N$ of constituent wave plates, provide higher order of stability against variations of the phase shift $\varphi$ and the light wavelength $\lambda$.
Examples of BB half-wave retarders are listed in Table \ref{Table} and their fidelities are illustrated in Fig.~\ref{Fig2}(top).

We can construct in the same manner various \emph{BB quarter-wave retarders}.
Their Jones matrix in the LR basis is
\begin{equation}
\J_{0}(\pm \pi /2) = \tfrac{1}{\sqrt{2}} \left[\begin{array}{cc}1 & \pm i \\ \pm i & 1 \end{array}\right].
\end{equation}
We have found that the most suitable composite sequence consists of $N-1$ $\lambda/2$ plates ($\varphi=\pi$) and a $\lambda/4$ plate ($\varphi=\pi/2$):
 $(\lambda/2)_{\theta_N}(\lambda/2)_{\theta_{N-1}}\cdots(\lambda/2)_{\theta_2}(\lambda/4)_{0}$ (read from right to left, cf.~Fig.~\ref{Fig1}(b)).
The corresponding Jones matrix is
\begin{equation}
\J^{(N)}=\J_{\theta_N}(\pi)\J_{\theta_{N-1}}(\pi)\cdots \J_{\theta_{2}}(\pi)\J_{0}(\pi/2).
\end{equation}
There are $N-1$ free phases $\theta_k$, with which we can nullify the first $\lfloor (N-1)/2 \rfloor$ complex derivatives
\begin{equation}
\left[ \partial _{\varphi }^{k}J_{12}^{\left( N\right) }\right] _{\varphi=\pi }=0\quad (k=1,2,\ldots, \lfloor (N-1)/2 \rfloor),
\label{nullify21}
\end{equation}
where $\lfloor x \rfloor$ denotes the integer part of $x$.
For even $N$ we can nullify also the real or imaginary part of the next non-zero derivative (of order $N/2$):
\begin{equation}
\text{Re}\left[\partial_{\varphi }^{N/2}\J_{12}^{\left( N\right) }\right] _{\varphi=\pi} = 0 \text{ or } \text{Im}\left[\partial_{\varphi }^{N/2}\J_{12}^{\left( N\right) }\right] _{\varphi=\pi} = 0.
\label{nullify22}
\end{equation}
Table \ref{Table} lists a set of phases that produce BB quarter-wave retarders; fidelities are shown in Fig.~\ref{Fig2}(bottom).

\section{Composite polarization retarders with a mirror}

We show now how to construct BB quarter-wave and half-wave retarders by placing a mirror at the end of the composite sequence of plates.
Thus the retarders are effectively twice as long and we can reduce the number of wave plates while maintaining the fidelity.
We note that if the incident light ``sees'' a wave plate rotated at an angle $\theta$ then the reflected light will ``see'' the same wave plate rotated at the angle $-\theta$.

For the BB \emph{half-wave retarder} the most efficient sequence is composed of $N-1$ half-wave plates ($\varphi=\pi$) and a quarter-wave plate ($\varphi=\pi/2$): $M(\lambda/4)_{\theta_N}(\lambda/2)_{\theta_{N-1}}\cdots(\lambda/2)_{\theta_2}(\lambda/2)_{\theta_1}$ (read from right to left, cf. Fig.~\ref{Fig1}(c)).
The total Jones matrix is
\begin{align}
\J^{(2N)} =& \J_{-\theta_{1}}(\pi) \J_{-\theta_{2}}(\pi)\cdots \J_{-\theta_{N-1}}(\pi) \J_{-\theta_N}(\pi/2) \bm{\sigma}_x \notag\\
 &\times \J_{\theta_{N}}(\pi/2) \J_{\theta_{N-1}}(\pi)\cdots \J_{\theta_{2}}(\pi) \J_{\theta_1}(\pi),
\end{align}
where the Pauli matrix $\bm{\sigma_x}$ describes the mirror in the LR basis.

First we impose $\J^{(2N)}=\J_0(\pi)$. We are left with $N-1$ rotation angles $\theta_k$, which we use to nullify the first $N-2$ (complex) derivatives
\begin{equation}
\left[ \partial _{\varphi }^{k}\J_{12}^{(2N) }\right] _{\varphi=\pi}=0 \quad ( k=1,2,\ldots, N-2)
\end{equation}
and the real or the imaginary part of the next non-zero derivative (of order $N-1$):
\begin{equation}
\text{Re}\left[\partial_{\varphi }^{N-1}\J_{12}^{\left( 2N\right) }\right] _{\varphi=\pi} = 0 \text{ or } \text{Im}\left[\partial_{\varphi }^{N-1}\J_{12}^{\left( 2N\right) }\right] _{\varphi=\pi} = 0.
\label{nullify22}
\end{equation}

For the BB \emph{quarter-wave retarder} the most suitable sequence is composed of $N-2$ half-wave plates ($\varphi=\pi$) surrounded by two quarter-wave plates ($\varphi=\pi/2$) as follows:
$M(\lambda/4)_{\theta_N}(\lambda/2)_{\theta_{N-1}}\cdots(\lambda/2)_{\theta_{2}}(\lambda/4)_{\theta_{1}}$ (read from right to left, cf. Fig.~\ref{Fig1}(d)).
The total Jones matrix is
\begin{align}
\J^{(2N)} =& \J_{-\theta_{1}}(\pi/2) \J_{-\theta_{2}}(\pi)\cdots \J_{-\theta_{N-1}}(\pi) \J_{-\theta_N}(\pi/2) \bm{\sigma}_x \notag\\
 &\times \J_{\theta_{N}}(\pi/2) \J_{\theta_{N-1}}(\pi)\cdots \J_{\theta_{2}}(\pi) \J_{\theta_1}(\pi/2).
\end{align}

\begin{figure}[t]
\centerline{\includegraphics[width=0.95\columnwidth]{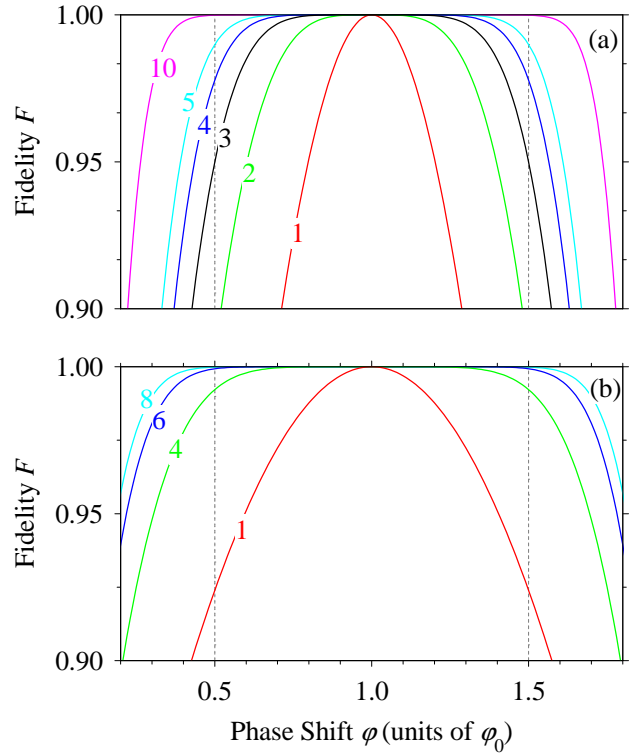}}
\caption{(Color online) Fidelity $F$ vs phase shift $\varphi$ for
 BB half-wave retarders with a mirror (frame (a) with $\varphi_0=\pi$, cf.~Fig.~\ref{Fig1}(c)) and
 BB quarter-wave retarders with a mirror (frame (b) with $\varphi_0=\pi/2$, cf.~Fig.~\ref{Fig1}(d)),
 for different number of constituent plates $N$.
The rotation angles are given in Table \ref{Table}.
}
\label{Fig3}
\end{figure}

We impose $\J^{(2N)}=\J_0(\pi/2)$, at the expense of two rotating angles $\theta_k$, and set the first $N-2$ (complex) derivatives to zero,
\begin{equation}
\left[ \partial _{\varphi }^{k}\J_{12}^{(2N) }\right] _{\varphi=\pi}=0 \quad ( k=1,2,\ldots,N-2 ).
\label{nullify the first 2n derivatives}
\end{equation}
Exemplary solutions are listed in Table \ref{Table}; fidelities are shown in Fig.~\ref{Fig3}.

\section{Conclusion}

The presented composite retarders are advantageous, as they allow to manipulate polarization only by varying the rotation angle of the single wave plates.
Therefore they offer robustness against variations of the parameters of both the crystal and the light field.
These include the crystal temperature, the wavelength of the electric field, the crystal length and the angle of incidence. In addition, the proposed retarders significantly outperform the existing BB retarders \cite{Harris64,Ardavan} for more than five individual plates. An experimental implementation with standard retarders available in most laboratories should be straightforward.

This work is supported by the European Commission project FASTQUAST, the Bulgarian NSF grants D002-90/08, DMU02-19/09 and Sofia University Grant 022/2011.


\end{document}